\documentclass[preprint]{aastex}
\usepackage{emulateapj5,apjfonts,epsfig}

\slugcomment{ApJ Letters, in press}
\shorttitle{BBN and Precision Cosmology}
\shortauthors{Burles, Nollett, and Turner}

\begin{document}
\newcounter{figs}

\title{Big-Bang Nucleosynthesis Predictions for Precision Cosmology}
\author{Scott Burles\altaffilmark{1,2},
Kenneth M. Nollett\altaffilmark{3,4}, and
Michael S. Turner\altaffilmark{2,3,5}}
\altaffiltext1{Experimental Astrophysics,
Fermi National Accelerator Laboratory,
Box 500, Batavia, IL~~60510-0500}
\altaffiltext2{Department of Astronomy and Astrophysics,
Enrico Fermi Institute, 5640 S. Ellis Avenue,
The University of Chicago, Chicago, IL~~60637-1433}
\altaffiltext3{Department of Physics, The University of Chicago,
Chicago, IL~~60637}
\altaffiltext4{Physics Department MS 130-33, California Institute of Technology,
Pasadena, CA~~91125}
\altaffiltext5{NASA/Fermilab Astrophysics Center,
Fermi National Accelerator Laboratory,
Box 500, Batavia, IL~~60510-0500}

\begin{abstract}

The determination of the primeval deuterium abundance has opened a
precision era in big-bang nucleosynthesis (BBN), making accurate
predictions more important than ever before.  We present in analytic
form new, more precise predictions for the light-element abundances
and their error matrix.  Using our predictions and the primeval deuterium 
abundance we infer a baryon density of $\Omega_Bh^2 = 0.020\pm 0.002$ (95\% cl)
and find no evidence for stellar production (or destruction)
of $^3$He beyond burning D to $^3$He.  Conclusions about
$^4$He and $^7$Li currently hinge upon possible systematic
error in their measurements.

\end{abstract}

\keywords{nuclear reactions, nucleosynthesis, abundances -- cosmology : theory }

\section{Introduction}

The BBN prediction of a large primeval abundance of $^4$He ($Y_P
\approx 0.25$) was the first success of the hot big-bang model.  For
two decades the consistency of the BBN predictions for the abundances
of D, $^3$He, $^4$He and $^7$Li with their inferred primeval
abundances has been an important test of the standard
cosmology at early times ($t\sim 1\,$sec).  BBN has also been used
to ``inventory'' ordinary matter at a simpler time and to probe
fundamental physics
(see e.g., Schramm \& Turner 1998 or Olive et al. 2000).

With the detection of the deuterium Ly-$\alpha$ feature in the
absorption spectra of 3 high-redshift ($z>2$) quasars and the accurate
determination of the primeval abundance of deuterium, (D/H)$_P=(3.0\pm
0.2)\times 10^{-5}$ (Burles et al. 2000;
Tytler et al. 2000; O'Meara et al. 2001)
BBN has entered a new precision era (Schramm \& Turner 1998).
Because its abundance depends strongly upon the baryon density, and its
subsequent chemical evolution is so simple (astrophysical processes
only destroy D), deuterium can accurately peg the baryon density.
Once determined, the baryon density allows the abundances of $^3$He,
$^4$He and $^7$Li to be predicted.  These predictions can be
used to test the consistency of the big-bang framework and
to probe astrophysics.

The chemical evolution of $^4$He is simple (stars produce it) and so
its predicted abundance, $Y_P = 0.246 \pm 0.001$ (Lopez \& Turner
1999), can be used as a consistency test of BBN and the standard
cosmology.  Because the $^7$Li abundance in old pop II stars may be
depleted, lithium probes both stellar models and the consistency of
the standard cosmology.

While the post-big-bang evolution of $^3$He is complex, the sum
D+$^3$He can be used to study the chemical evolution of the Galaxy.
All stars burn D to $^3$He as they evolve onto the main sequence (MS).
Later stages of stellar evolution may produce or destroy $^3$He,
depending on stellar mass and subject to uncertainty in modeling.
Thus, the evolution of D+$^3$He measures the net stellar production of
$^3$He beyond pre-MS burning (Yang et al. 1984), providing an
important probe of stellar models.

A key to realizing BBN's full potential in the precision era is
accurate and reliable predictions.  With this in mind, we have
recently used Monte-Carlo techniques (Burles et al. 1999; Nollett \&
Burles 2000) to link the calculated abundances directly to the nuclear
data, making the predictions more reliable.

Previous work (Smith et al. 1993, Fiorentini et al. 1998) was based on
fitting cross-section data to standard forms, estimating conservative
uncertainties to accommodate most or all of the data for each
reaction.  Very recent work (Esposito et al. 2000; Vangioni-Flam et
al. 2000) computes only ``maximum'' uncertainties, using upper and
lower limits quoted in a compilation of charged-particle reaction
rates (Angulo et al. 1999).  In contrast, 
our procedure ties the abundance errors directly
to the experimental measurements by weighting the data by
their quoted errors, and furthermore, leads to smaller estimated
abundance errors (by factors of 2-3).

In this {\em Letter} we present our results in the form of analytic
fits for the abundances and their error matrix. We then use these
predictions to make
inferences about the baryon density, the consistency of BBN, $^7$Li
depletion and stellar $^3$He production.

\section{Analytic Results}

Our BBN code draws reaction rates from a statistical distribution and
computes the corresponding distribution of BBN yields.  It varies all
of the laboratory data (over 1200 individual data points)
simultaneously, drawing random realizations of each data point and
normalizations for each data set from Gaussian distributions
representing reported values and uncertainties.  For each realization,
the BBN yields are computed using thermally-averaged smooth
representations of the realized data.  The results presented here are
based on 25 000 such realizations of the data (see Fig. 1).  More
details are given in Burles et al (1999), and Nollett \& Burles
(2000).

Here we present fits of the means, variances, and correlation matrix
of the predicted abundances to fifth-order polynomials in $x\equiv
\log_{10}\eta +10$, where $\eta$ is the baryon-to-photon ratio; see
Tables 1--3.  Applicable over the range $0\le x \le 1$, our fits are
accurate to better than 0.2\% for the abundances and 10\% for the
variances.  For the mean $^4$He yields, we adopt the fitting formula
of Lopez \& Turner (1999)
which is accurate to 0.05\%\footnote[6]{Our fit coefficients are the
$a_i$ from Eq. 44.  We note that their fitting
formula for the dependence of $Y_P$ on neutron lifetime ($\delta Y_P$
in Equation 43 of that paper) has a misprint in the signs but not the
magnitudes of the coefficients $b_i$.  The correct sequence of signs
for the $b_i$ is $++-+-$.  They also provide a fit for the
$N_\nu$ dependence of $Y_P$.}.  Some of the abundances follow approximate
power laws, and so we have obtained accurate fits by fitting the means
and variances of their base-ten logarithms in all cases except the
mean $Y_P$.  Because our estimates for the uncertainties are small,
${\rm Var}(Y_i) = ({\bar Y}_i/0.4343)^2\,{\rm Var}(\log_{10}Y_i)$.
The covariance matrix is written in terms of the variances and a
correlation matrix $r_{ij}$:
\begin{equation}
\rho_{ij} = r_{ij}\sqrt{{\rm Var}(Y_i){\rm Var}(Y_j)}.
\end{equation}
where $Y_i$ is baryon fraction for $^4$He and number relative to
Hydrogen for the other nuclides, and $\bar{Y}_i$ is its mean over the
output yields.

Finally, because BBN produces $^7$Li by two distinct processes, direct
production and indirect production through $^7$Be with subsequent
electron capture to $^7$Li, we have split the $^7$Li yield into these
two pieces to obtain more accurate fits.  The mean prediction for
$^7$Li is just the sum of the two contributions; the variance ${\rm
Var}(Y_7)={\rm Var}(Y_{\rm Li}) + {\rm Var}(Y_{\rm Be}) + 2\rho_{\rm
Li,Be}$.  The covariance between the total BBN $^7$Li and another
nuclide $\rho_{i,7} = \rho_{i,{\rm Li}} + \rho_{i,{\rm Be}}$.

\begin{table*}
\caption{Fits to the abundances, $\log_{10}\bar{Y}_i=\sum_i a_i x^i$,
except, $\bar{Y}_P=\sum_i a_i x^i$.}
\label{tab:yields}
\begin{tabular}{lcccccc}
Nuclide & $a_0$ & $a_1$ & $a_2$ & $a_3$ & $a_4$ & $a_5$ \\
\tableline
$Y_P$       &  0.22292 & 0.05547 & -0.05639 & 0.04587 & -0.01501 & --\\
D/H         &  -3.3287 &  -1.6277 &  -0.2286 &   0.7794 &  -0.6207 & 0.0846 \\
$^3$He/H    &  -4.4411 &  -0.7955 &   0.4153 &  -0.9909 &   1.0925 & -0.3924 \\
(D+$^3$He)/H & -3.2964 &  -1.5675 &  -0.1355 &   0.8018 &  -0.7421 & 0.2225 \\
$^7$Li/H    & -9.2729 &  -2.1707 &  -0.6159 &   4.1289 &  -3.6407 &  0.7504 \\
$^7$Be/H    & -12.0558 &   3.6027 &   2.7657 &  -6.5512 &   4.4725 & -1.1700 \\
\end{tabular}
\end{table*}

\begin{table*}
\caption{Fits to the variances,
${\rm Var}(\log_{10}Y_i)=\sum_i a_i x^i$.}
\label{tab:variances}
\begin{tabular}{lcccccc}
Nuclide & $a_0$ & $a_1$ & $a_2$ & $a_3$ & $a_4$ & $a_5$ \\
\tableline
$10^5\,Y_P$  & 0.2544 &  -1.3463 &   4.0384 &  -6.3448 &   4.9910 &  -1.5446 \\
$10^3\,$(D/H) &  0.2560 &  0.1379 & -2.3363 &  5.0495 & -4.6972 &  1.9351 \\
$10^3\,(^3$He/H) & 0.0776 &  0.1826 & -0.7725 &  1.5357 & -0.9106 &  0.1522 \\
$10^3\,$(D+$^3$He)/H & 0.2181 & -0.0287 & -1.6284 &  3.5182 & -2.8499 &  0.8323 \\
$10^2\,(^7$Li/H) & 0.2154 & -0.0049 & -1.7200 &  4.0635 & -3.8618 &  1.3946 \\
$10^2\,(^7$Be/H) & 0.7970 &  1.2036 & -6.5462 &  6.0483 & -0.2788 & -1.1190 \\
\end{tabular}
\end{table*}

\begin{table*}
\caption{Fits to the correlation coefficients,
$r_{j,k}=\sum_i a_i x^i$.}
\label{tab:coeff}
\begin{tabular}{cccccccc}
\multicolumn{2}{c}{Coefficient $j,k$} &  $a_0$ & $a_1$ & $a_2$ & $a_3$ & $a_4$ & $a_5$ \\
\tableline
  $Y_P$     &     D      & -0.8121 &   0.6430 &   3.3284 &  -7.2925 &   5.6748 &  -1.5914 \\
  $Y_P$     &  $^3$He    &  0.2129 &   1.3468 &  -8.3646 &  15.8093 & -12.8939 &   3.9055 \\
  $Y_P$     & D + $^3$He & -0.8091 &   0.6468 &   3.3848 &  -7.4565 &   5.7605 &  -1.5838 \\
  $Y_P$     &   $^7$Li   & -0.3630 &  -0.1017 &   5.1531 & -10.3563 &   7.5445 &  -1.8680 \\
  $Y_P$     &   $^7$Be   &  0.7744 &  -0.3414 &  -4.0492 &   8.4836 &  -6.7167 &   1.9345 \\
     D      &  $^3$He    & -0.1924 &  -1.9722 &   8.2683 & -13.6301 &   8.1108 &  -1.2999 \\
     D      & D + $^3$He &  0.9995 &  -0.0238 &   0.1229 &  -0.2574 &  -0.1625 &   0.1352 \\
     D      &   $^7$Li   &  0.4219 &   0.2824 &  -0.9063 &  -6.9928 &  14.5503 &  -6.8278 \\
     D      &   $^7$Be   & -0.8820 &  -0.0647 &  -0.4330 &   3.9867 &  -4.9394 &   1.6666 \\
  $^3$He    & D + $^3$He & -0.1526 &  -1.7701 &   7.2981 &  -9.4669 &   3.7557 &   0.1560 \\
  $^3$He    &   $^7$Li   & -0.1321 &  -0.8465 &   3.1187 &   0.7518 &  -6.1419 &   2.9935 \\
  $^3$He    &   $^7$Be   &  0.3293 &   1.6390 &  -6.3839 &   8.9361 &  -4.2279 &   0.3574 \\
 D + $^3$He &   $^7$Li   &  0.4186 &   0.3165 &  -1.2759 &  -6.0646 &  14.4155 &  -7.2783 \\
 D + $^3$He &   $^7$Be   & -0.8744 &  -0.0455 &  -0.3596 &   3.9249 &  -4.6197 &   1.5773 \\
   $^7$Li   &   $^7$Be   & -0.4091 &  -0.1971 &  -0.5008 &  11.8943 & -19.0115 &   8.0258 \\
\end{tabular}
\end{table*}

{
\refstepcounter{figs}
\label{fig:abundances}
        \centerline{{\vbox{\epsfxsize=8cm\epsfbox{etaplot.eps}}}}
{\footnotesize{FIG. 1.  Predicted big-bang abundances of the light
elements shown as bands of 95\% confidence.  }}
\bigskip
}

\section{Implications}

To use our predictions we need observed abundances of the
light elements.  This is a lively area of research, with some controversy.
Here, based upon our evaluation of the data,
we state our choices with brief justification
and point the reader interested in more detail to the relevant
literature.

For the primordial deuterium abundance we use the
weighted average of the 3 detections
in high-redshift Ly-$\alpha$,
(D/H)$_P=(3.0\pm 0.2)\times 10^{-5}$
(for further discussion see Burles et al. 2000;
Tytler et al. 2000; O'Meara et al. 2001).

For the present abundance of D+$^3$He, we use measurements of both
elements made in the local interstellar medium (ISM).  The deuterium
abundance, D/H$=(1.5\pm 0.2 \pm 0.5)\times 10^{-5}$, comes from HST,
IUE and Copernicus measurements along 12 lines of sight to nearby
stars (Linsky 1998; Lemoine et al. 1999; McCullough 1992).  The first
error is statistical, and the second error represents the possibility
of scatter due to spatial variations (Vidal-Madjar \& Gry 1984; Linsky
1998; Vidal-Madjar et al. 1999; Sonneborn et al. 2000); as it turns
out, the uncertainty in $^3$He dominates both.  Gloeckler \& Geiss
(1998) have determined the ratio of $^3$He to $^4$He in the local ISM
using the pick-up ion technique.  Allowing for a local $^4$He mass
fraction between 25\% and 30\%, their measurement translates to
$^3$He/H$=(2.2\pm 0.8)\times 10^{-5}$ and (D+$^3$He)/H$=(3.7\pm
1)\times 10^{-5}$.

For the primordial $^7$Li abundance we use the value advocated by Ryan
(2000), based upon the extant measurements of $^7$Li in the
atmospheres of old halo stars.  His value,
$^7$Li/H$=1.2^{+0.35}_{-0.2} \times 10^{-10}$, includes empirical
corrections for cosmic-ray production, stellar depletion, and improved
atmospheric models, and the uncertainty arises mainly from these
corrections.  This is consistent with other estimates (see e.g.,
Bonifacio \& Molaro 1997; Ryan et al. 1999; Thorburn 1994).

The primordial abundance of $^4$He is best inferred from H{\sc ii}
regions in metal-poor, dwarf emission-line galaxies.  While such
measurements are some of the most precise in astrophysics, the values
for $Y_P$ obtained from the two largest samples of such objects are
not consistent and concerns remain about systematic error.

Olive et al (1997) have compiled a large sample of objects and find
$Y_P = 0.234 \pm 0.002$.  On the other hand, Izotov \& Thuan (1998)
have assembled a large sample from a single observational program,
extracting $Y_P$ from the spectra by a different method.  They find
$Y_P =0.244\pm 0.002$ (consistent with the earlier sample of Kunth \&
Sargent 1983, which found $Y_P = 0.245\pm 0.003$).  Further, they have
shown that at least one of the most metal-poor objects (IZw18) used in
the earlier sample suffered from stellar absorption, and argue that it
and possibly other metal-poor objects in this sample explain the
discrepancy.  Viegas et al. (2000) argue that the Izotov and Thuan
sample should be corrected downward by a small amount ($\Delta Y_P
\approx 0.003$) to account for neutral and doubly ionized $^4$He;
Ballantyne et al.  (2000) agree on the magnitude of the effect, but
not the direction.  Finally, a recent study of different parts of a
single H{\sc ii} region in the SMC finds $Y= 0.241 \pm 0.002$
(Peimbert et al. 2000), at face value implying $Y_P \le 0.241\pm
0.002$.

Clearly, the final word on $Y_P$ is not in.  For now, because of the
homogeneity and size of the Izotov and Thuan sample and the possible
corruption of the other sample by stellar absorption, with caution we
adopt $Y_P=0.244\pm 0.002$.  (Had we adopted an intermediate value,
with a systematic error reflecting the discrepancy between the two
data sets, our conclusions would be largely the same.)

Using these choices, we have constructed separate likelihood functions
for the baryon-to-photon ratio $\eta$ from the abundances of D,
D+$^3$He, $^4$He and $^7$Li, assuming Gaussian distributions for the
uncertainties; see Fig.~2.  While the D, D+$^3$He
and $^4$He abundances are all consistent with $\eta \approx 5\times
10^{-10}$, most precisely pegged by D, the $^7$Li abundance favors a
significantly lower value.  Combining these, we find $\chi^2 = 23.2$
for 3 degrees of freedom (4 abundances minus 1 parameter).  This is
the well-known lithium problem: the deuterium-inferred value for the
baryon density predicts a $^7$Li abundance that is about $3 \sigma$
larger than that measured in old pop II halo stars (see e.g., Burles
et al. 1999; or Olive et al. 2000)

Since it is possible, and some stellar models suggest, that there has
been more depletion of $^7$Li in old halo stars than the 5\% inferred
by Ryan (2000), we introduce a model parameter, $f_7$, the ratio of
the inferred $^7$Li/H in old pop II stars to its predicted primordial
value.  It quantifies how much the primordial $^7$Li/H has been
affected by additional stellar depletion, cosmic-ray production, or
theoretical difficulties (e.g., systematic errors in the nuclear cross
sections or in the modeling of stellar atmospheres).  An $f_7 \neq 1$
might also reflect fundamental problems, such as systematic problems
with the deuterium abundance or inconsistencies in BBN.

{
\refstepcounter{figs}
\label{fig:face-value}
        \centerline{{\vbox{\epsfxsize=8cm\epsfbox{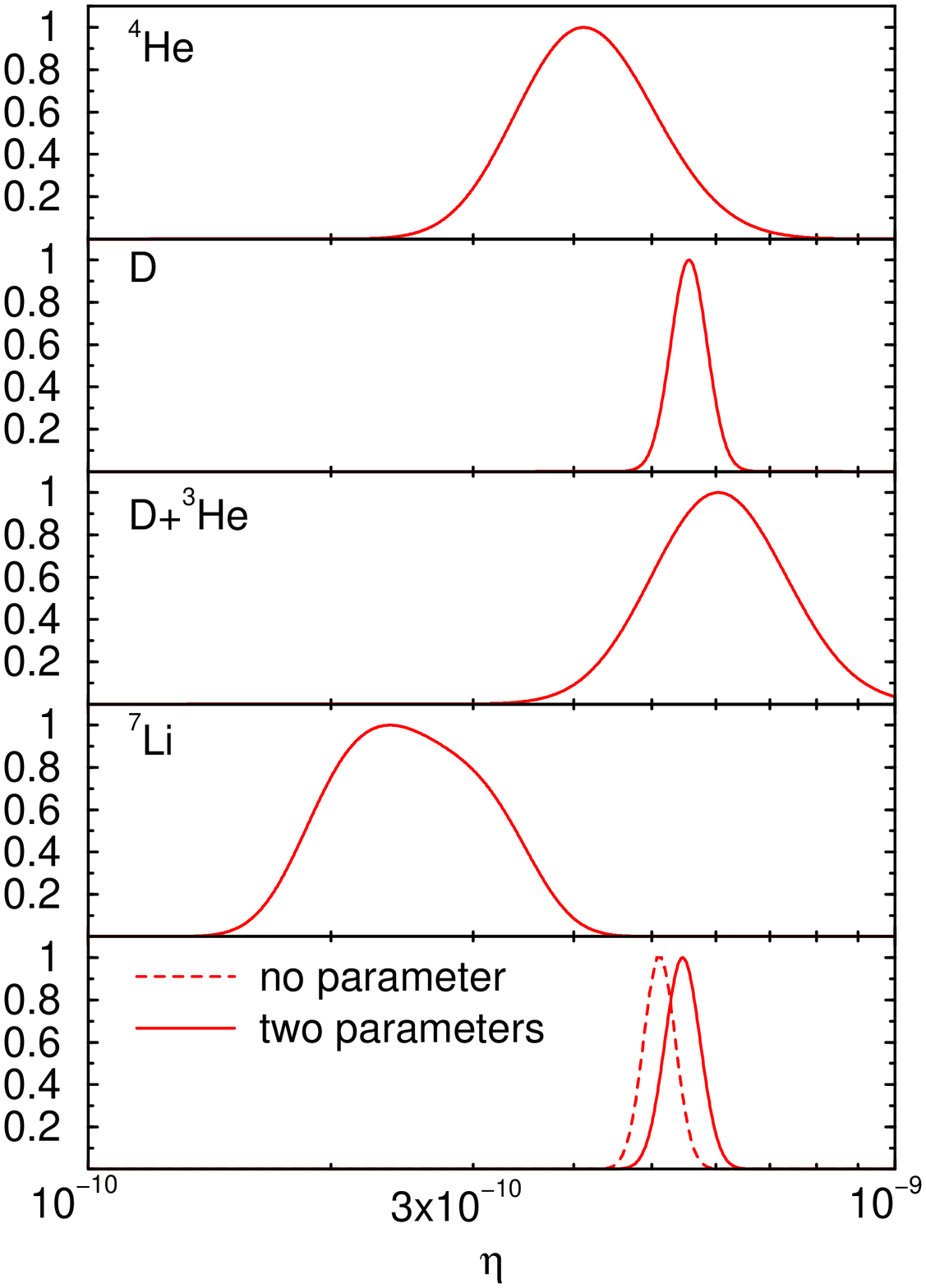}}}}
{\footnotesize{FIG. 2.  Likelihood functions, normalized to unit
maximum, for $\eta$ derived from single-abundance analyses (top four
panels) and from simultaneous analysis of all four abundances and
their covariances (bottom panel).  In the ``two-parameter'' analysis
the likelihood is marginalized over $f_7$ and $f_{23}$ (see text).  }}
\smallskip
}

Since there is no reason to believe that the present value of $({\rm
D}+ ^3{\rm He})/{\rm H}$ in the local ISM is the primordial value, we
introduce an analogous factor, $f_{23}$, which is the ratio of the
present $({\rm D}+ ^3{\rm He})/{\rm H}$ to its primordial value.  If
$f_{23}=1$, then the light-element abundances are consistent with the
simple hypothesis that stars only convert D into $^3$He during pre-MS
burning, conserving D+$^3$He by number; $f_{23}> 1$ indicates
additional net stellar production of $^3$He, and $f_{23}<1$ indicates
net stellar destruction of $^3$He after the pre-MS.

Fig.~3 shows the distributions for $f_7$ and $f_{23}$,
each marginalized over our other two parameters (e.g., the $f_{23}$
curve results from marginalizing over $\eta$ and $f_7$).  The most
likely value for $f_7$ is 0.32, with 95\% confidence interval
$0.20-0.55$.  That is, consistency between the deuterium-predicted
$^7$Li abundance and the pop II abundance requires a depletion of
greater than a factor of two or some as-yet unidentified source of
systematic error in the BBN prediction or measurement.  Such depletion
can be achieved in stellar models and still be consistent with other
observational constraints, including the plateau in $^7$Li abundance
in old pop II stars and the detection of $^6$Li in several stars (see
e.g., Vauclair \& Charbonnel 1995; Pinsonneault et al. 1999; 
Salaris \& Weiss 2001).

The most likely value for $f_{23}$ is 0.88, with 95\% confidence
interval $0.55-1.54$.  Unlike $f_7$, this new parameter has
essentially no effect on the question of concordance, and its value
supports the simple hypothesis of only pre-MS $^3$He production.  It
also disfavors stellar models that predict significant net $^3$He
production (or destruction) and is consistent with an earlier
comparison of pre-solar and ISM measurements of D+$^3$He which showed
no evidence for an increase over the last 4.5\,Gyr (Turner et
al. 1996).  This is somewhat surprising, since the conventional models
for the galactic chemical evolution of $^3$He predict a significant
increase in D+$^3$He due to net $^3$He production by low mass stars
(Iben \& Truran 1978; Dearborn, Schramm, \& Steigman 1986).  However,
Wasserburg, Boothroyd, \& Sackmann (1995) argue that $^3$He
destruction by some low-mass stars is possible.

\vskip 10pt

{
\refstepcounter{figs}
\label{fig:d23}
        \centerline{{\vbox{\epsfysize=7.2cm\epsfbox{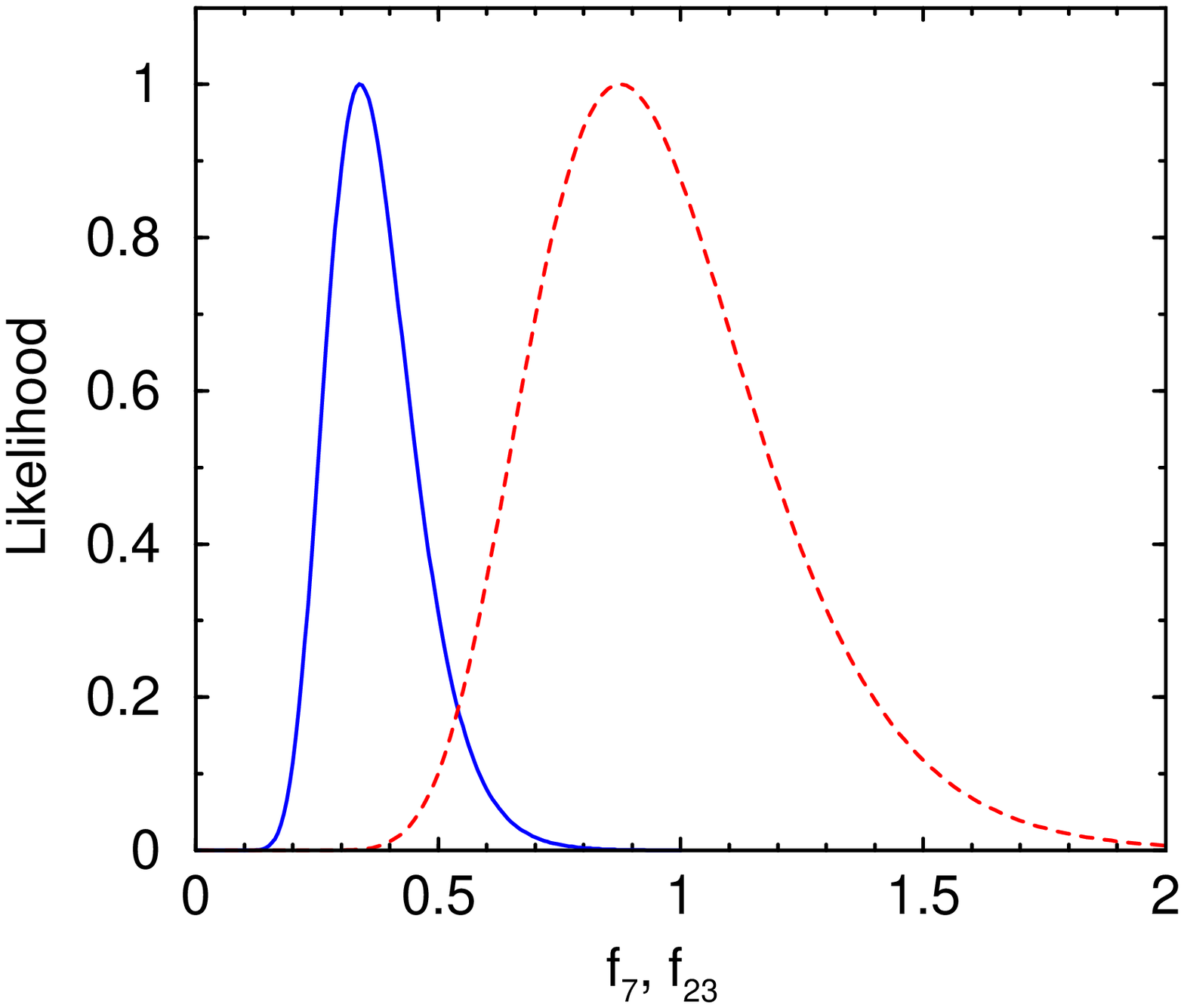}}}}
{\footnotesize{FIG. 3.  Marginalized likelihoods for $f_7$
(solid curve) and $f_{23}$ (broken curve).}}
\bigskip
}

BBN and the light-element abundances can be used to fix the
baryon-to-photon ratio at the end of BBN ($t\sim 200\,$sec).  We have
computed the likelihood function for the baryon density using the
abundances of all four light elements and marginalizing over $f_7$ and
$f_{23}$, giving all the weight to D and $^4$He.  We find: $\eta =
(5.5 \pm 0.5)\times 10^{-10}$, shown as the ``two-parameter'' curve in
Fig.~2.  The value is driven almost entirely by
deuterium: using the deuterium alone we find $\eta = (5.6\pm
0.5)\times 10^{-10}$.

To relate $\eta$ to the present
baryon density ($\Omega_Bh^2$) one needs to know the present photon
temperature ($T=2.725\,{\rm K}\pm 0.001\,$K) and the average mass per
baryon number, $\bar m = 1.6700\ (1.6701)\times 10^{ -24}\,$g for the post-BBN
mix (solar abundance), and make the assumption that the expansion
has been adiabatic since BBN.  Then, $\eta$ and the baryon density are
related by $\Omega_Bh^2 = (3.650\pm 0.004) \times 10^7 \,\eta_{\rm BBN}$.
Within the standard cosmology, $\eta = (5.5\pm 0.5)\times 10^{-10}$
translates to a baryon density,
\begin{equation}
\Omega_Bh^2 = 0.020\pm 0.002 \qquad {\rm (95\%\ confidence)}.
\end{equation}

Finally, we mention other recent likelihood analyses of BBN.  Hata et
al. (1995) addressed the consistency of BBN, focusing especially on
$^4$He and $^7$Li (also see Copi et al. 1995).  Olive \& Thomas (1997)
and Fiorentini et al. (1998) carried out assessments of BBN using
older estimates of the theoretical errors and a broader range for
primordial D/H.  The present analysis is the first using the new
Nollett \& Burles (2000) error estimates as well as the recently
clarified primeval D/H.

\section{Concluding Remarks}

We have presented analytical fits for our new, more accurate
predictions of the light-element abundances and their error matrix.
These results have already renewed interest in more accurately
determining key nuclear rates, and further improvements are likely
(see e.g., Schreiber et al. 2000; Rupak 2000).

Using our results and the primeval deuterium abundance
from three high-redshift Ly-$\alpha$ systems we infer
$\Omega_Bh^2 = 0.020\pm 0.002$ (95\% cl).
For $h=0.7\pm 0.07$, this implies a baryon fraction
$\Omega_B = 0.041\pm 0.009$, with the error dominated
by that in $H_0$.
Measurements of cosmic microwave background (CMB) anisotropy
have recently also determined the baryon density.  The
physics underlying this method is very different -- gravity-driven
acoustic oscillations in the Universe at 500 000 yrs -- but the result
is similar: $\Omega_Bh^2 = 0.032^{+0.009}_{-0.008}$ at 95\% cl (Jaffe
et al. 2000).  While there is about a $2\sigma$ difference, this first
result from CMB anisotropy confirms the long-standing BBN prediction
of a low baryon density, and with it, the need for nonbaryonic dark matter.

The sum of D+$^3$He predicted for the deuterium baryon density is
consistent with that in the ISM today, implying no significant
production (or destruction) of $^3$He beyond pre-MS burning.  The
deuterium-predicted $^4$He abundance is an important consistency test
of BBN; however, systematic measurement error (the $Y_P$ values from
the two key data sets differ by 5 times the statistical error)
precludes firm conclusions at this time.  Likewise, the discrepancy
between the predicted $^7$Li abundance and the abundance measured in
pop II stars has no simple explanation: The discrepancy could indicate
that $^7$Li has been depleted in pop II stars by about a factor of
two, that uncertainties (in observations or predictions) have been
grossly underestimated, or that there is an inconsistency in BBN.


\end{document}